\documentclass[aps,preprint]{revtex4}%
\usepackage{amsfonts}
\usepackage{amsmath}
\usepackage{amssymb}
\usepackage{graphicx}%
\providecommand{\U}[1]{\protect\rule{.1in}{.1in}}

\begin{document}
\preprint{HEP/123-qed}
\title[Short title for running header]{Quantum Gravity: physics from supergeometries}
\author{D.J. Cirilo-Lombardo }
\affiliation{Bogoliubov Laboratory of Theoretical Physics Joint Institute for Nuclear
Research 141980, Dubna (Moscow region), Russian Federation}
\author{Thiago Prudencio}
\affiliation{Coordination of Interdisciplinary Course in Science and Technology, Federal
University of Maranhao - UFMA, Campus Bacanga, 65080-805, Sao Luis-MA, Brazil.}
\affiliation{Department of Physics, Federal University of Maranhao - UFMA, Campus Bacanga,
65085-580 Sao Luis-MA, Brazil.\}}
\pacs{PACS number}

\begin{abstract}
We show that the metric (line element) is the first geometrical object to be
associated to a discrete (quantum) structure of the spacetime without
necessity of black hole-entropy-area arguments, in sharp contrast with other
attempts in the literature. To this end, an emergent metric solution obtained
previously \ in [Physics Letters B 661, 186-191 (2008)] from a particular
non-degenerate Riemmanian superspace is introduced. This emergent metric is
described by a physical coherent state belonging to the metaplectic group
$Mp\left(  n\right)  $ with a Poissonian distribution at lower $n$ (number
basis) restoring the classical thermal continuum behaviour at large $n$
($n\rightarrow\infty)$, or leading to non-classical radiation states, as is
conjectured in a quite general basis by mean the Bekenstein-Mukhanov effect.
Group-dependent conditions that control the behavior of \ the macroscopic
regime spectrum (thermal or not), as the relationship with the problem of area
/ entropy of the black hole are presented and discussed.

\end{abstract}
\volumeyear{year}
\volumenumber{number}
\issuenumber{number}
\eid{identifier}
\date[Date text]{date}
\received[Received text]{date}

\revised[Revised text]{date}

\accepted[Accepted text]{date}

\published[Published text]{date}

\startpage{101}
\endpage{102}
\maketitle
\tableofcontents

\section{Introduction}

The unification of gravity and quantum theory is one of the great challenges
of physics. The last years were dominated by attempts to reach this goal by
rather radical new concepts, as is exemplified by the string theory and loop
quantum gravity. Then, one of the main points treated in the current
literature is the close relation between the quantum structure of the
spacetime and its discretization at particular scale. In the pioneering works
of the last century, the concept of fundamental scale was associated to the
minimal length or, geometrically speaking, to the metric (e.g.through the line
element describing the spacetime). Actually, contrarily to these prior
investigations, arguments favoring the use of the area as a fundamental entity
were appearing in recent years. This fact was motivated strongly by the
relation area / entropy of the black hole on one side, and the theoretical
structure of theories such as loop quantum gravity (spin networks), dynamical
triangulations and however, the string theory on the other side. Two questions
motivated by the gravity-quantum unification immediately arise. One of them
without answer until today: is the exact behaviour at macroscopic regime of
the quantum gravity thermal or not? The another question is: there exists a
consistent formulation where the length is the minimum fundamental entity and
that the spectrum of such theory meets the correct limits? (e.g. correct
spacing between levels for n small and for large n). We will demonstrate
through this letter that such consistent description certainly exists, and it
can contemplate the classical (thermal) or not, behaviour of the spectrum. Our
claim is based in some previous research of one of us trying to give an
unambiguous quantum mechanical description of a particle in a general
spacetime. Because the introduction of supersymmetry provide new approach to
several physical problems of interest, in [1] a particular interesting
Riemannian $N=1$ superspace was introduced. The main feature of this
superspace that makes it especially important, is that the corresponding
supermetric, which is the basic ingredient of a Volkov-Pashnev particle action
[8], is \emph{invertible and non-degenerate}, that is, of $G4$ type in the
Casalbuoni's classification [10] As shown in [2,3,5], the non-degeneracy of
the supermetrics (and therefore of the corresponding superspaces) leads to
important consequences in the description of physical systems [5]. In
particular, notorious geometrical and topological effects on the quantum
states, namely, \textit{consistent mechanisms of localization and
confinement}, due purely to the supergeometrical character of the Lagrangian.
Also \textit{an alternative to the Randall-Sundrum (RS) model without extra
bosonic coordinates}, can be consistently formulated in terms of such non
degenerated superspace approach, eliminating the problems that the RS-like
models present at the quantum level [1,2,3]. And from the probabilistic point
of view was recently demonstrated in [4] that, using the probability current
as the probability density, the quantum counterpart of the Fisher's metric can
be exactly implemented being all the relevant quantum operators exactly
constructed in a manner that was already inferred in 1988 on a quite general
basis by Caianiello. In this letter (strongly motivated by the above results)
we will show, after brief description of the superspace and the emergent
spacetime of ref.[1,2,3], that as a result of quantization of this
supersystem[2,3], a discrete metric associated with coherent states (in
obedience to a Poisson distribution), is immediately obtained without
prescription of discretization. This discrete solution, that represents an
emergent metric, is described by a physical coherent state belonging to the
metaplectic group $Mp\left(  n\right)  $ with a Poissonian distribution at
lower n (number basis) restoring the classical thermal continuum behaviour at
large $n(n\rightarrow\infty,$ number basis$)$, or leading to non-classical
radiation states, as is conjectured in a quite general form by the dynamic
Bekenstein-Mukhanov effect. The results that we present here are absolutely
without black/hole entropy arguments given a priory. Finally a discussion
linking our results with the black hole entropy and spectrum are giving and
some perspectives and future directions of research suggested.

\section{Supermetric and emergent spacetime}

The model introduced in [1,8], represents a free particle in a superspace with
coordinates $z_{A}\equiv\left(  x_{\mu},\theta_{\alpha},\overline{\theta
}_{\overset{\cdot}{\alpha}}\right)  $. It is described by the Lagrangian
density
\begin{equation}
\mathcal{L}=-m\sqrt{\omega^{A}\omega_{A}}=-m\sqrt{\overset{\circ}{\omega}%
_{\mu}\overset{\circ}{\omega}^{\,\mu}+{\mathbf{a}}\dot{\theta}^{\alpha}%
\dot{\theta}_{\alpha}-\mathbf{a}^{\ast}\dot{\bar{\theta}}^{\dot{\alpha}}%
\dot{\bar{\theta}}_{\dot{\alpha}}}. \label{CSL}%
\end{equation}
where $\overset{\circ}{\omega_{\mu}}=\overset{.}{x}_{\mu}-i(\overset{.}%
{\theta}\ \sigma_{\mu}\overline{\theta}-\theta\ \sigma_{\mu}\overset
{.}{\overline{\theta}})$, and the dot indicates derivative with respect to the
parameter $\tau$, as usual. In coordinates, the line element of the superspace
reads,
\begin{equation}
ds^{2}=\dot{z}^{A}\dot{z}_{A}=\dot{x}^{\mu}\dot{x}_{\mu}-2i\dot{x}^{\mu}%
(\dot{\theta}\sigma_{\mu}\bar{\theta}-\theta\sigma_{\mu}\dot{\bar{\theta}%
})+\left(  {\mathbf{a-}}\bar{\theta}^{\dot{\alpha}}\bar{\theta}_{\dot{\alpha}%
}\right)  \dot{\theta}^{\alpha}\dot{\theta}_{\alpha}-\left(  {\mathbf{a}%
}^{\ast}+\theta^{\alpha}\theta_{\alpha}\right)  \dot{\bar{\theta}}%
^{\dot{\alpha}}\dot{\bar{\theta}}_{\dot{\alpha}}\nonumber
\end{equation}

Is important to note that the quantization was exactly performed by a new
method introduced by one of us in [2,3] given the correct physical and
mathematical interpretation to the square root Hamiltonian. The method is
based on two fundamental points: first, introducing a modification of Lanczos
technique [2] that permits to pick the correct phase space of the problem
without modify the form of the relevant quantum geometrical operators (i.e.the
particular form of the Hamiltonian or Lagrangian remains as square root ). And
second, using the underlying covering group of $SL(2C)$ ( that is the
Metaplectic group) to give a quantum meaning to the radical operator
(Lagrangian or Hamiltonian). With these ingredients the problem is
schematically reduced to $\mathcal{H}\left\vert \Psi\right\rangle \equiv
\sqrt{m^{2}-\mathcal{P}_{0}\mathcal{P}^{0}-\left(  \mathcal{P}_{i}%
\mathcal{P}^{i}+\frac{1}{a}\Pi^{\alpha}\Pi_{\alpha}-\frac{1}{a^{\ast}}%
\Pi^{\overset{.}{\alpha}}\Pi_{\overset{.}{\alpha}}\right)  }\left\vert
\Psi\right\rangle =0$ were $\mathcal{P}^{\mu},$ $\Pi^{\alpha}$ are the momenta
corresponding to the supercoordinates.

Without lose generality and for simplicity, the `squared' solution with three
compactified dimensions ( $\lambda=2$ spin fixed) is [1,3,5]
\begin{equation}
g_{AB}(t)=e^{A(t)+\xi\varrho(t)}g_{AB}(0), \label{CSsol}%
\end{equation}
where the initial values of the metric components are given by
\begin{equation}
g_{ab}(0)=\langle\psi(0)|\left(
\begin{array}
[c]{c}%
a\\
a^{\dagger}%
\end{array}
\right)  _{ab}|\psi(0)\rangle,
\end{equation}
or, explicitly,
\begin{align}
g_{\mu\nu}(0)  &  =\eta_{\mu\nu}\,,\qquad g_{\mu\alpha}(0)=-i\sigma_{\mu
\alpha\dot{\alpha}}\bar{\theta}^{\dot{\alpha}}\,,\qquad g_{\mu\dot{\alpha}%
}(0)=-i\theta^{\alpha}\sigma_{\mu\alpha\dot{\alpha}}\,,\\
g_{\alpha\beta}(0)  &  =(a-\bar{\theta}^{\dot{\alpha}}\bar{\theta}%
_{\dot{\alpha}})\epsilon_{\alpha\beta}\,,\qquad g_{\dot{\alpha}\dot{\beta}%
}(0)=-(a^{\ast}+\theta^{\alpha}\theta_{\alpha})\epsilon_{\dot{\alpha}%
\dot{\beta}}\,. \label{gdiego}%
\end{align}
It worth mention here that these components were obtained in the simplest case
in [9].

The bosonic and spinorial parts of the exponent in the superfield solution
(\ref{CSsol}) are, respectively,
\begin{equation}%
\begin{array}
[c]{rcl}%
A(t) & = & -\left(  \frac{m}{|{\mathbf{a}}|}\right)  ^{2}t^{2}+c_{1}t+c_{2},\\
\xi\varrho\left(  t\right)  & = & \xi\left(  \phi_{\alpha}(t)+\bar{\chi}%
_{\dot{\alpha}}(t)\right) \\
& = & \theta^{\alpha}\left(  \overset{\circ}{\phi}_{\alpha}\cos\left(  \omega
t/2\right)  +\frac{2}{\omega}Z_{\alpha}\right)  -\overline{\theta}%
^{\overset{\cdot}{\alpha}}\left(  -\overset{\circ}{\overline{\phi}}%
_{\overset{\cdot}{\alpha}}\sin\left(  \omega t/2\right)  -\frac{2}{\omega
}\overline{Z}_{\overset{.}{\alpha}}\right) \\
& = & \theta^{\alpha}\overset{\circ}{\phi}_{\alpha}\cos\left(  \omega
t/2\right)  +\bar{\theta}^{\overset{\cdot}{\alpha}}\overset{\circ}%
{\overline{\phi}}_{\overset{\cdot}{\alpha}}\sin\left(  \omega t/2\right)
+4|\mathbf{a}|Re(\theta Z),
\end{array}
\label{expo}%
\end{equation}
where $\overset{\circ}{\phi}_{\alpha},Z_{\alpha},\overline{Z}_{\overset
{.}{\beta}}$ are constant spinors,
$\omega=1/|\mathbf{a}|$ and the constant $c_{1}\in\mathbb{C}$, due to the
obvious physical reasons and the chiral restoration limit of the superfield
solution [1,3,5].

\section{Superspace and discrete spacetime structure}

Now we will see how the discrete spacetime structure naturally arise from the
model,. Expanding on a number basis, as usual
\[
\sum_{m}|m\rangle\langle m|=1,
\]
we have
\[
g_{ab}(0)=\sum_{n,m}\langle\psi(0)|m\rangle\langle m|L_{ab}|n\rangle\langle
n|\psi(0)\rangle
\]
then
\[
g_{ab}(t)=\underset{f\left(  t\right)  }{\underbrace{e^{A(t)+\xi\rho(t)}}}%
\sum_{n,m}\langle\psi(0)|m\rangle\langle n|\psi(0)\rangle\langle
m|L_{ab}|n\rangle
\]%
\begin{equation}
\langle m|L_{ab}|n\rangle=\langle m|\left(
\begin{array}
[c]{c}%
a\\
a^{\dagger}%
\end{array}
\right)  _{ab}|n\rangle=\left(
\begin{array}
[c]{c}%
\langle m|n-1\rangle\sqrt{n}\\
\langle m|n+1\rangle\sqrt{n+1}%
\end{array}
\right)  _{ab}=\left(
\begin{array}
[c]{c}%
\delta_{m,n-1}\sqrt{m}\\
\delta_{m,n+1}\sqrt{m+1}%
\end{array}
\right)  _{ab} \tag{7}%
\end{equation}
It follows%
\[
g_{ab}(0)=\sum_{n,m}\langle\psi(0)|m\rangle\left(
\begin{array}
[c]{c}%
\delta_{m,n-1}\sqrt{m}\\
\delta_{m,n+1}\sqrt{m+1}%
\end{array}
\right)  _{ab}\langle n|\psi(0)\rangle
\]%
\[
g_{ab}(0)=\sum_{n}\sqrt{n}\langle\psi(0)|n-1\rangle\langle n|\psi
(0)\rangle\left(
\begin{array}
[c]{c}%
1\\
0
\end{array}
\right)  _{ab}+\sum_{m}\sqrt{n+1}\langle\psi(0)|n+1\rangle\langle
n|\psi(0)\rangle\left(
\begin{array}
[c]{c}%
0\\
1
\end{array}
\right)  _{ab}%
\]
From the equation above we see that the only clear sense for it is due the
decomposition of $\psi$ into the basic states of the metapletic
representation
\begin{equation}
|\psi(0)\rangle=A|\alpha_{+}\rangle+B|\alpha_{-}\rangle\tag{8}%
\end{equation}
where the constants $A$ and $B$ are arbitrary and they control the classical
behavior of the spectrum at the macroscopic level. We, without lose generality
in this part of the discussion, take $A=B$ such that $|\psi(0)\rangle
=|\alpha_{+}\rangle+|\alpha_{-}\rangle,$ but we will return to this important
point later. This, in fact, is the effect of the decomposition of the SO(2,1)
group in two irreducible representations of the metaplectic group Mp(2):
spanning even and odd n respectively. The important feature of the
\ state$|\psi(0)\rangle=|\alpha_{+}\rangle+|\alpha_{-}\rangle$ is that is
invariant (if $A=B$) to the action of operators $a$ and $a^{\dagger}$. This
fact is because in the metaplectic representation the general behaviour of
these states are: $a|\alpha_{+}\rangle=a^{\dag}|\alpha_{+}\rangle=|\alpha
_{-}\rangle$ and $a|\alpha_{-}\rangle=a^{\dag}|\alpha_{-}\rangle=|\alpha
_{+}\rangle$(we will not enter in more details here).

Is easily checked from the Poissonian distribution for the coherent states:
$P_{\alpha}(n)=|\langle n|\alpha\rangle|^{2}=\frac{\alpha^{n}e^{-\alpha}}{n!}$
obeying $\underset{n=0}{\overset{\infty}{\sum}}P_{\alpha}(n)=1,$
\ \ \ \ $\underset{n=0}{\overset{\infty}{\sum}n}P_{\alpha}(n)=\alpha$\ \ that
it differs with the individual distributions coming from each one of the two
irreducible representations of the metaplectic group Mp(2) (spanning even and
odd n respectively):.
\begin{equation}
\underset{n=0}{\overset{\infty}{\sum}}P_{\alpha_{+}}(2n)=e^{-\alpha}%
\cosh(\alpha),\text{ \ \ }\underset{n=0}{\overset{\infty}{\sum}}P_{\alpha_{-}%
}(2n+1)=e^{-\alpha}\sinh(\alpha)\rightarrow\underset{n=0}{\overset{\infty
}{\sum}}\left(  P_{\alpha_{+}}(n)+P_{\alpha_{-}}(n)\right)  =1 \tag{9}%
\end{equation}
Although the different form between above equations, the limit \ n$\rightarrow
\infty$ is the same for the sum of the two distributions coming from the
Mp$\left(  2\right)  $ irreducible representations (Irreps). and for the
SO(2,1) representation as it should be.

Having this in mind, the specific form of $|\alpha_{+}\rangle,|\alpha
_{-}\rangle$ was given in [1,2,3] and are
\begin{align}
|\alpha_{+}\rangle &  \equiv\left\vert \Psi_{1/4}\left(  0,\xi,q\right)
\right\rangle =\overset{+\infty}{\underset{k=0}{\sum}}f_{2k}\left(
0,\xi\right)  \left\vert 2k\right\rangle =\overset{+\infty}{\underset
{k=0}{\sum}}f_{2k}\left(  0,\xi\right)  \frac{\left(  a^{\dag}\right)  ^{2k}%
}{\sqrt{\left(  2k\right)  !}}\left\vert 0\right\rangle \nonumber\\
|\alpha_{-}\rangle &  \equiv\left\vert \Psi_{3/4}\left(  0,\xi,q\right)
\right\rangle =\overset{+\infty}{\underset{k=0}{\sum}}f_{2k+1}\left(
0,\xi\right)  \left\vert 2k+1\right\rangle =\overset{+\infty}{\underset
{k=0}{\sum}}f_{2k+1}\left(  0,\xi\right)  \frac{\left(  a^{\dagger}\right)
^{2k+1}}{\sqrt{\left(  2k+1\right)  !}}\left\vert 0\right\rangle \tag{10}%
\end{align}
where in the parameter $\xi$ all the possible remaining $B_{1}$ (odd)
dependence is stored. Then, we arrive to following result {\small
\begin{gather}
g_{ab}(t)=\frac{f\left(  t\right)  }{2}\sum_{m}\left\{  \left[  P_{\alpha_{+}%
}(2m)\cdot2m+P_{\alpha-}(2m+1)\cdot\left(  2m+1\right)  \right]  \left(
\begin{array}
[c]{c}%
1\\
0
\end{array}
\right)  _{ab}+\right. \nonumber\\
+\left.  \left[  P_{\alpha_{+}^{\ast}}(2m)\cdot2m+P_{\alpha_{-}^{\ast}%
}(2m+1)\cdot(2m+1)\right]  \left(
\begin{array}
[c]{c}%
0\\
1
\end{array}
\right)  _{ab}\right\}  \tag{11}%
\end{gather}
}this expression is the core of our discussion: it shows explicitly the
discrete structure of the spacetime as the fundamental basis for a consistent
quantum field theory of gravity. By the other hand, when we reach the limit
n$\rightarrow\infty$ the metric solution goes to the continuum due:
$\underset{n=0}{\overset{\infty}{\sum}}\left[  P_{\alpha_{+}}(2m)\cdot
2m+P_{\alpha-}(2m+1)\cdot(2m+1)\right]  =\underset{n=0}{\overset{\infty}{\sum
}}\left[  P_{\alpha_{+}}(2m+2)\cdot\left(  2m+2\right)  +P_{\alpha_{-}%
}(2m+1)\cdot(2m+1)\right]  =\alpha e^{-|\alpha|}\left(  \cosh(\alpha
)+\sinh(\alpha)\right)  =\alpha$ and similarly for the lower part (spinor
down) of above equation $\underset{n=0}{\overset{\infty}{\sum}}\left[
P_{\alpha_{+}}(2m)\cdot2m+P_{\alpha_{-}}(2m+1)\cdot(2m+1)\right]
=\alpha^{\ast}.$ Consequently, when the number of levels increase the metric
solution goes to the continuum "manifold" general relativistic behaviour:%
\begin{equation}
g_{ab}(t)_{n\rightarrow\infty}\rightarrow\frac{f\left(  t\right)  }{2}\left\{
\alpha\left(
\begin{array}
[c]{c}%
1\\
0
\end{array}
\right)  _{ab}+\alpha^{\ast}\left(
\begin{array}
[c]{c}%
0\\
1
\end{array}
\right)  _{ab}\right\}  =f\left(  t\right)  \langle\psi(0)|\left(
\begin{array}
[c]{c}%
a\\
a^{\dagger}%
\end{array}
\right)  _{ab}|\psi(0)\rangle\tag{12}%
\end{equation}
as expected.

\section{The minimal length}

Is not difficult to see that for $m=0$ the metric solution takes its minimal value

{\small
\begin{align}
g_{ab}(t)  &  =\frac{f\left(  t\right)  }{2}\left[  P_{\alpha-}(1)\left(
\begin{array}
[c]{c}%
1\\
0
\end{array}
\right)  _{ab}+P_{\alpha_{-}^{\ast}}(1)\left(
\begin{array}
[c]{c}%
0\\
1
\end{array}
\right)  _{ab}\right] \nonumber\\
&  =\frac{f\left(  t\right)  }{2}e^{-\left\vert \alpha\right\vert }\left[
\alpha\left(
\begin{array}
[c]{c}%
1\\
0
\end{array}
\right)  _{ab}+\alpha^{\ast}\left(
\begin{array}
[c]{c}%
0\\
1
\end{array}
\right)  _{ab}\right]  \tag{13}%
\end{align}
}this evidently defines the minimal length due the metric axioms in a
Riemannian manifold. This point will be not analyzed fully here, but in
principle (due the existence of discrete Poincare subgroups of this
supermetric) fundamental symmetries as the Lorentz one, can be preserved at
this level of discretization. Some of interested issues related to this
problem in general were given in ref. [11].

\section{Black hole entropy and superspace solution}

As is well known, the black hole entropy $S=k_{B}A_{bh}/4l_{P}^{2}$ where is
the horizon area and $l_{P}\equiv\sqrt{\hslash G/c^{3}}$ is the Planck length
was first found by Bekenstein and Hawking [18] using thermodynamic arguments
of preservation of the first and second laws of thermodynamics. An information
theory proof was also found by Bekenstein in which can be treated as the
measure of "inaccessibility" of the information of an external observer on an
actual internal configuration of the black hole realized in a given state
(described by values of mass, charge, and angular momentum)". The first
controversial thing that immediately appears from the point of view of
statistical mechanics (in which the entropy is the mean logarithm of the
density matrix) is that the entropy of a black hole is proportional to its
surface area. About this issue, Bekenstein propose a model of quantization of
the horizon area in the section with the suggestive title "Demystifying black
hole's entropy proportionality to area" in ref.[19]. Resuming the proposal,
the horizon is formed by patches of equal area $\delta l_{P}^{2}$ (however,
which are added one after another at a time). Their standard size is important
and makes them all equivalent. The horizon can be regarded as having many
degrees of freedom, one per each patch, due it is made from equivalent patches
all with the same number $\chi$ of quantum states. Consequently, the total
number of quantum states of the horizon is $\Omega_{H}=\chi^{A_{bh}/\delta
l_{P}^{2}}$ and the statistical (Boltzmann) entropy associated with the
horizon is $S=k_{B}\ln\Omega_{H}=k_{B}\left(  A_{bh}/\delta l_{P}^{2}\right)
\ln\chi$: putting $\delta=4\ln\chi$ Bekenstein arrived to the expected
thermodynamical black hole formula. As was pointed out in [20], Bekenstein
don't gives account that putting $\delta$ into the original black hole entropy
formula we obtain the Poisson expression for the total number of states,
\begin{equation}
\Omega_{H}=e^{A_{bh}/4l_{P}^{2}} \tag{14}%
\end{equation}
being this precisely \textit{the link} with the structure of the emergent
coherent state metric. Considering the similar Poissonian expression for the
number of states from $g_{ab}$, namely $e^{\left\vert \alpha\right\vert }$,
the relation between the coherent state eigenvalue $\alpha$ corresponding to
the solution (11) and the above equation is clear:%
\begin{equation}
A_{bh}/4l_{P}^{2}=\left\vert \alpha\right\vert \tag{15}%
\end{equation}
this expression relates the space of phase of the coherent state solution
$g_{ab}$ and the $A_{bh}$ through $\ $the $l_{P}^{2}$ and $\left\vert
\alpha\right\vert .$ The linear behaviour of area and entropy with respect
to$\alpha$ is given in Figure 1 (this important issue will be focused by the
authors in a separate work).%

\begin{figure}
[ptb]
\begin{center}
\fbox{\includegraphics[
height=3.0787in,
width=5.7in
]%
{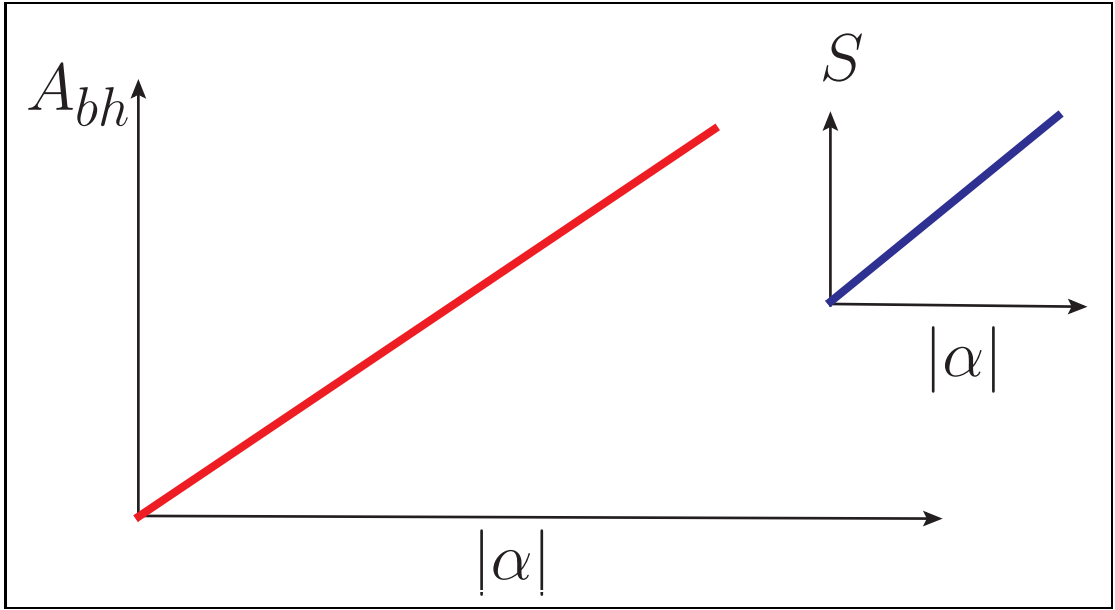}%
}\caption{ Linear behavior of A$_{bh}$ and entropy with $\left\vert
\alpha\right\vert $}%
\end{center}
\end{figure}

\section{Is the black hole radiation black?}

Recently was discussed the fact if the black hole Hawking's radiation is
finally thermal or it can be quantically affected, as suggested again by
Bekenstein and Mukhanov [13]. Due the interplay between the area of the black
hole surface and the black hole mass, it is likely to be quantized as well.
Then, the mass of the black hole decreases when radiation is emitted due the
quantum jump from one quantized value of the mass (energy) to a lower
quantized value. In consequence (because radiation is emitted at quantized
frequencies, corresponding to the differences between energy levels) quantum
gravity implies a discretized emission spectrum for the black hole radiation.

As is well known from the comments of the (LQG)\ loop quantum gravity
community [12,14,16,17], the spectral lines can be very dense in macroscopic
regimes leading physically no contradiction with Hawking's prediction of a
continuous thermal spectrum. Was also verified that the behavior of the
spectrum is ansatz-dependent at macroscopic regimes: if we pick (as Bekenstein
and Mukhanov does) the simplest ansatz for the quantization of the area -that
the area is quantized in multiple integers of an elementary area $A_{0}$, then
the emitted spectrum turns out to be macroscopically discrete, this effect as
the kinematical Bekenstein-Mukhanov effect. By the other hand, into the loop
quantum gravity context, the celebrated thermal spectrum is reached because
the density of levels increase in parallel with the number of levels. Then, is
possible to circumvented the theoretical dilemma?: from the point of view of
LQG the spectrum is always continuous at macroscopic regime, and from the
viewpoint of Mukhanov and Bekenstein the spectrum may be macroscopically
discrete (but ansatz-dependent finally). From the point of view of our model
we can give an affirmative answer to this question: if now we suppose simply
that $A\neq B$ in the state solution (8,11), that is one possibility in our
proposal due the arbitrariness of the constants A and B, we have
$|\psi(0)\rangle=A|\alpha_{+}\rangle+B|\alpha_{-}\rangle$ then we cannot reach
the thermal (Hawking) spectrum at the macroscopic level. This fact is clear
because we need exact balance between the superposition of the two irreducible
representations of the Metaplectic group as clearly given by expressions
(11,9). This will leads, as a result, non classical states of radiation in the
sense of [21] as can be easily seen putting, for example, $B$ $\left(
\text{or }A\right)  $ zero:%
\begin{equation}
g_{ab}(t)=A\frac{f\left(  t\right)  }{2}\sum_{m}\left[  P_{\alpha_{+}%
}(2m)\cdot\left(  2m\right)  +P_{\alpha-}(2m+1)\cdot(2m+1)\right]  \left(
\begin{array}
[c]{c}%
1\\
0
\end{array}
\right)  _{ab} \tag{16}%
\end{equation}
notice that only the up spinor part survives and the classical (thermal) limit
is not reached, even in the continuous limit were the number of levels
increases accordingly to%
\begin{equation}
g_{ab}(t)_{n\rightarrow\infty}\rightarrow\frac{f\left(  t\right)  }{2}%
A\alpha\left(
\begin{array}
[c]{c}%
1\\
0
\end{array}
\right)  _{ab}=Af\left(  t\right)  \langle\psi(0)|\left(
\begin{array}
[c]{c}%
a\\
0
\end{array}
\right)  _{ab}|\psi(0)\rangle\tag{17}%
\end{equation}
In such a case $A=0$ $(B=0)$, the spectrum will takes only even(odd) levels
becoming evidently non thermal. Then, if $A=B$ the kinematical
Bekenstein-Mukhanov effect disappears and the thermal Hawking spectrum is
reached at the continuum classical gravity level (the Poissonian behaviour of
the distribution is complete). Otherwise, with $A\neq B$, the spectrum belongs
to a non classical one and the quantum properties of the gravity are
macroscopically manifest.

\section{Concluding remarks}

Trough this paper we have been shown that a N=1 superspace equipped with a
nondegenerate and invertible supermetric where the unconstrained quantization
was exactly performed by new methods based on coherent states and respecting
the form of the Hamiltonian (modified Lanczos technique), a discrete structure
of spacetime naturally emerges without any prescription of discretization (in
sharp contrast of the other attempts in the literature) . Due the Metaplectic
representation (double covering of the $SL(2C)$) of the coherent state
solution representing the emergent spacetime, the crossover from the quantum
to the macroscopical regime (classical or not) is natural and consistent. This
important fact permits us to conciliate the two apparently different pictures
of the macroscopical quantum gravity regime given by the LQG\ claims
supporting the Hawking (Thermal) spectrum) and the dynamical
Bekenstein-Mukhanov effect that point out that quantum (non thermal) imprints
can survive at macroscopical regime. Despite the simplicity of the model
introduced here, we have been obtained physically and geometrically, an amount
of important answers with respect to a consistent quantum gravity formulation.
The main properties that any consistent formulation of quantum gravity must
have, in the light of the results presented, are:

1) Emergent nature of the spacetime.

2) Independence of the discretization method.

3) Consistent suitable transition to the macroscopic (classical,
semiclassical, etc.) regime.

4) Total and absolute independence of particular solutions or other arguments
involving particular geometries (e.g. black-hole/area and the entropy ) .

5) Solutions, arguments involving particular geometries, etc. of the previous
point, must be reached by the quantum gravity theory but not depending them at
the fundamental level.

\section{Acknowlegements}

DJCL\ is very grateful to the JINR\ Directorate and the BLTP for his
hospitality and financial support.

\section{References}

[1] Diego Julio Cirilo-Lombardo; \textit{The geometrical properties of
Riemannian superspaces, exact solutions and the mechanism of localization}.
Physics Letters \textbf{B} 661,(2008) 186-191.

[2] Diego Julio Cirilo-Lombardo; \textit{Non-compact groups, Coherent States,
Relativistic Wave equations and the Harmonic Oscillator}. Foundations of
Physics 37 (2007) 919-950.

[3] Diego Julio Cirilo-Lombardo; \textit{Non-compact Groups, Coherent States,
Relativistic Wave Equations and the Harmonic Oscillator II: Physical and
geometrical considerations.} Found Phys 39 (2009) 373--396.

[4] Diego Julio Cirilo-Lombardo with V.I. Afonso, \textit{Information metric
from Riemannian superspaces}. Phys.Lett\textbf{.A}376 (2012) 3599 .

[5] Diego Julio Cirilo-Lombardo; \textit{Geometrical properties of Riemannian
superspaces, observables and physical states}, The European Physical Journal C
- Particles and Fields, (2012), Volume 72, Number 7, 2079

[6] See for example: M. B. Green, J.H. Schwartz and E. Witten,
\textit{Superstring Theory I and II}, (CUP, Cambridge 1988)

[7] C Rovelli, L Smolin, \textit{Knot Theory and Quantum Gravity} Phys Rev
Lett 61 (1988)1155, \textit{Loop space representation of quantum general
relativity}, Nucl Phys \textbf{B}331 (1990) 80. For various perspectives on
loop quantum

gravity, see: A Ashtekar, in \textit{Gravitation and Quantization}, Les Houches

1992, edited by B Julia and J Zinn-Justin (Elsvier Science: Paris 1995).; L
Smolin, in \textit{Quantum Gravity and Cosmology},

ed J Perez-Mercader, J Sola, E Verdaguer (World Scientific: Singapore 1992).

[8] D.V. Volkov, A.I. Pashnev,\textit{\textquotedblleft Supersymmetric
lagrangian for particles in proper time\textquotedblright, }Theoret. and Math.
Phys. 44 (3) (1980) 770.

[9] V.P. Akulov, D.V. Volkov, \textit{Riemannian superspaces of minimal
dimensionality,} Theoret. and Math. Phys. 41 (2) (1979) 939.

[10] R. Casalbuoni, \textit{Relativity and Supersymmetries, }Phys. Lett.
\textbf{B} 62 (1976) 49.

[11] L. J. Garay: \textquotedblleft Quantum Gravity and Minimal
Length\textquotedblright, Int.J.Mod.Phys. \textbf{A}10 (1995) 145-166

[12] See also: C Rovelli: \textit{\textquotedblleft Black Hole Entropy from
Loop Quantum Gravity\textquotedblright}, [gr-qc/9603063].

[13] J.D. Bekenstein, VF Mukhanov, \textit{\textquotedblleft Spectroscopy of
the quantum black hole\textquotedblright}, [gr-qc/9505012].

[14] L Smolin: \textit{\textquotedblleft Microscopic Deviations from Hawking
radiation?\textquotedblright}, Matters of Gravity 7, [gr-qc/9602001].

[15] LJ Garay: \textquotedblleft\textit{Quantum Gravity and Minimal
Length\textquotedblright}, Int.J.Mod.Phys. A10 (1995) 145-166

[16] A Ashtekar, C Rovelli, L Smolin, \textit{Weaving a classical metric with
quantum threads}, Phys Rev Lett 69 (1992) 237

[17] C Rovelli, L Smolin, \textit{Discreteness of area and volume in quantum
gravity,} Nucl. Phys. \textbf{B} 442 (1995) 593; R DePietri, C Rovelli,
\textit{\textquotedblleft Geometry Eigenvalues and Scalar Product}

\textit{from Recoupling Theory in Loop Quantum Gravity\textquotedblright},
Phys.Rev. \textbf{D} 54 (1996), [gr-qc/9602023]; A Ashtekar, J Lewandowski:
\textquotedblleft Quantum Theory of

Geometry I: Area Operator\textquotedblright, [gr-qc/9602046].

[18] J.D Bekenstein, \textit{Black Holes and Entropy}, Phys Rev \textbf{D}7
(1973) 2333. SW Hawking, \textit{Black hole explosions}, Nature 248 (1974) 30.

[19] J.D Bekenstein. Quantum black holes as atoms,(1997) [gr-qc/9710076] .

[20] A. G. Bashkirov and A. D. Sukhanov, \textit{Entropy of open quantum
systems and the Poisson distribution}, Theoretical and Mathematical Physics,
Vol. 123, No. 1, (2000), 504.

[21] V V Dodonov, \textit{`Nonclassical' states in quantum optics: a
`squeezed' review of the first 75 years, }J. Opt. B: Quantum Semiclass. Opt.
(2002)4 R1 .

\end{document}